\documentclass[12pt]{article}
\usepackage{epsf}
\usepackage{graphicx}
\usepackage{array,dcolumn}      % belongs to LaTeX2e tools
\usepackage{axodraw}
\usepackage{amssymb}
\newcommand{\hs}{\hspace*{0.5cm}}

\newcommand{\be}{\begin{equation}}
\newcommand{\ee}{\end{equation}}
\newcommand{\bea}{\begin{eqnarray}}
\newcommand{\eea}{\end{eqnarray}}
\newcommand{\nn}{\nonumber}

\newcommand{\la}{\lambda}

\newcommand{\ga}{\gamma}

\newcommand{\fr}{\frac}
\newcommand{\bc}{\begin{center}}
\newcommand{\ec}{\end{center}}

\newcommand{\si}{\sigma}

\begin{document}

\bc {\Large Self-interacting dark matter and
 Higgs bosons \\ in the
 $\mbox{SU}(3)_C\otimes \mbox{SU}(3)_L \otimes \mbox{U}(1)_N$
 model\\ with right-handed neutrinos}\\
\vspace*{1cm}

{\bf Hoang Ngoc Long$^{a,b}$ and Nguyen Quynh Lan$^{a}$}\\
\vspace*{0.5cm}

$^a$ {\it  Institute of  Physics, NCST,
P. O. Box 429, Bo Ho, Hanoi 10000, Vietnam}\\

$^b$ {\it The Abdus Salam International Centre for Theoretical
Physics, Trieste 34014, Italy.} \vspace*{1cm}

\ec

\begin{abstract}
We investigate the possibility that dark matter could be made from
$CP$-even and $CP$- odd Higgs bosons in the $\mbox{SU}(3)_C\otimes
\mbox{SU}(3)_L \otimes \mbox{U}(1)_N$ (3-3-1) model with
right-handed neutrinos. This self-interacting dark matters
 are stable without imposing of new symmetry and should be
weak-interacting.
\end{abstract}

PACS number(s): 95.35.+d, 12.60.Fr, 14.80.Cp

\noindent \hs It is an amazing fact that even as our understanding
of cosmology progresses by leaps and bounds, we remain almost
completely ignorant about the nature of most of the  matter in the
universe~\cite{mt00}. Cosmological models with a mixture of
roughly 35\% collisionless cold dark matter such as axions, WIMPs,
or any other candidate interacting through the weak and
gravitational forces only, and 65\% vacuum energy or quintessence
match observation of the cosmic microwave background and large
scale structure on extra-galactic scales with remarkable
accuracy~\cite{bah99,wan00}. It is known that only a fraction of
the dark matter can be made of ordinary baryons and its enormous
amount has unknown, nonbaryonic origin~\cite{pos01}. The nature of
dark matter is still a challenging question in cosmology.

\hs Until a few years ago, the more satisfactory cosmological
scenarios were those ones composed of ordinary matter, cold dark
matter and a contribution associated with the cosmological
constant. To be consistent with inflationary cosmology, the
spectrum of density fluctuations would be nearly scale-invariant
and adiabatic. However, in recent years it has been pointed out
that the conventional models of collisionless cold dark matter
lead to problems with regard to galactic structures. They were
only able to fit the observations on large scales ($\gg 1$ Mpc).
Also, $N$-body simulations in these models result in a central
singularity of the galactic halos~\cite{ghi} with a large number
of sub-halos~\cite{moore}, which are in conflict with astronomical
observations. A number of other inconsistencies are discussed in
Refs.~\cite{dav01,bul00}. Thus, the cold dark matter model is not
able to explain observations on scales smaller than a few Mpc.

\hs However, it has recently been shown that an elegant way to
avoid these problems is to assume the so called {\it
self-interacting dark matter}~\cite{ss}. One should notice that,
in spite of all, self-interacting models lead to spherical halo
centers in clusters, which is not in agreement with ellipsoidal
centers indicated by strong gravitational lens
observations~\cite{yoh00} and by Chandra
observations~\cite{buote}.

However, self-interacting dark matter models are self-motivated as
alternative models.  It is a well-accepted fact that the plausible
candidates for dark matter are elementary particles. The key
property of these particles is that,  they  must  have a large
scattering  cross-section and negligible annihilation or
dissipation. The Spergel-Steinhard  model  has motivated many
follow-up studies~\cite{pos01,ost,mcd}. Several authors have
proposed models in which a specific scalar singlet that satisfies
the self-interacting dark matter properties is introduced in the
standard model (SM) in an {\it ad hoc} way \cite{pos01,mcd}.

\hs The SM offers no options for dark matter. The first gauge
model for  SIDM were found by Fregolente and Tonasse~\cite{ft} in
the  3-3-1 model. It is to be noted that in the model considered
in~\cite{ft}
 to keep the Higgs sector with three triplets
one has to propose an existence of  {\it exotic leptons}. The
3-3-1 models were proposed with an independent
motivation~\cite{ppf}. These models have the following intriguing
features such as
 the models are anomaly free only if the number
of families $N$ is a multiple of three. If further one adds the
condition of QCD asymptotic freedom, which is valid only if the
number of families of quarks is to be less than five, it follows
that $N$ is equal to 3.

\hs A subject that has not been given much attention by particle
physicists in the past, could prove to be a remarkable powerful
and precise probe of the properties of dark matter.

\hs The aim of this paper is to show that
 the   3-3-1 model with right-handed (RH)
neutrinos~\cite{flt} contains  such self-interacting dark matter.

\hs To frame the context, it is appropriate to recall briefly some
relevant features of the 3 - 3 - 1 model with RH
neutrinos~\cite{flt}.
 In this model the leptons are in triplets, and the third member
is a RH neutrino: \be f^{a}_L = \left(
               \nu^a_L, e^a_L, (\nu^c_L)^a
\right)^T \sim (1, 3, -1/3), e^a_R\sim (1, 1, -1). \label{l2} \ee

The first two generations of quarks are in antitriplets while the
third one is in a triplet: \be Q_{iL} = \left(
                d_{iL},-u_{iL}, D_{iL}\\
                 \right)^T \sim (3, \bar{3}, 0),
\label{q} \ee
\[ u_{iR}\sim (3, 1, 2/3), d_{iR}\sim (3, 1, -1/3),
D_{iR}\sim (3, 1, -1/3),\ i=1,2,\] \be
 Q_{3L} = \left(
                 u_{3L}, d_{3L}, T_{L}
                 \right)^T \sim (3, 3, 1/3),
\ee
\[ u_{3R}\sim (3, 1, 2/3), d_{3R}\sim (3, 1, -1/3), T_{R}
\sim (3, 1, 2/3).\] The charged gauge bosons are defined as \bea
\sqrt{2}\ W^+_\mu &=& W^1_\mu - iW^2_\mu ,
\sqrt{2}\ Y^-_\mu = W^6_\mu - iW^7_\mu ,\nn\\
\sqrt{2}\ X_\mu^o &=& W^4_\mu - iW^5_\mu. \eea

\hs  The {\it physical} neutral gauge bosons are again related to
$Z, Z'$ through the mixing angle $\phi$.

The symmetry breaking can be achieved with just three
$\mbox{SU}(3)_L$ triplets

\bea \chi & = &\left(
                \chi^0, \chi^-, \chi^{,0}
                 \right)^T \sim (1, 3, -1/3),\nn\\
\rho & =& \left(
                \rho^+, \rho^0, \rho^{,+}
                  \right)^T \sim (1, 3, 2/3),\label{hig}\\
\eta & =& \left(
                \eta^0, \eta^-, \eta^{,0}
                 \right)^T \sim (1, 3, -1/3),\nn.
\eea The necessary VEVs are \be \langle\chi \rangle = (0, 0,
\omega/\sqrt{2})^T,\ \langle\rho \rangle = (0, u/\sqrt{2}, 0)^T,\
\langle\eta \rangle = (v/\sqrt{2}, 0, 0)^T. \label{vev} \ee

 After symmetry
breaking the gauge bosons gain  masses \be
m^2_W=\frac{1}{4}g^2(u^2+v^2),\
M^2_Y=\frac{1}{4}g^2(v^2+\omega^2),
M^2_X=\frac{1}{4}g^2(u^2+\omega^2). \label{rhb} \ee
Eqn.(\ref{rhb}) gives us a relation \be v_W^2 = u^2 + v^2 =246^2\
\  \mbox{GeV}^2. \label{vw} \ee \hs  In order to be consistent
with the low energy phenomenology we have to assume that $\langle
\chi \rangle \gg\ \langle \rho \rangle,\ \langle \eta \rangle$
such that $m_W \ll M_X, M_Y$.

\hs  The symmetry-breaking hierarchy gives us splitting on the
bilepton masses~\cite{til} \be | M_X^2 - M_Y^2 | \leq m_W^2.
\label{mar} \ee

\hs Our aim in this paper is to show that the 3-3-1 model with RH
neutrinos furnishes a good candidate for (self-interacting) dark
matter. The main properties that a good dark matter candidate must
satisfy are stability and neutrality. Therefore, we go to the
scalar sector of the model, more specifically to the neutral
scalars, and we examine whether any of them can be stable and in
addition whether they can satisfy the self-interacting dark matter
criterions \cite{ss}. In addition, one should notice that such
dark matter particle must not overpopulate the Universe. On the
other hand, since our dark matter particle is not imposed
arbitrarily to solve this specific problem, we must check that the
necessary values of the parameters do not spoil the other bounds
of the model.

\hs Under assumption of the discrete symmetry  $\chi \rightarrow -
\chi$, the most general potential can then be written in the
following form~\cite{l97} \bea V(\eta,\rho,\chi)&=&\mu^2_1 \eta^+
\eta +
 \mu^2_2 \rho^+ \rho +  \mu^2_3 \chi^+ \chi +
\lambda_1 (\eta^+ \eta)^2 + \lambda_2 (\rho^+ \rho)^2 +
\lambda_3 (\chi^+ \chi)^2 \nonumber \\
& + & (\eta^+ \eta) [ \lambda_4 (\rho^+ \rho) + \lambda_5 (\chi^+
\chi)] + \lambda_6 (\rho^+ \rho)(\chi^+ \chi) +
\lambda_7 (\rho^+ \eta)(\eta^+ \rho) \nonumber\\
& + & \lambda_8 (\chi^+ \eta)(\eta^+ \chi) + \lambda_9 (\rho^+
\chi)(\chi^+ \rho) +
 \lambda_{10} (\chi^+ \eta + \eta^+ \chi)^2.
\label{pot} \eea We rewrite the expansion of the scalar fields
which acquire a VEV: \be \eta^o = \fr{1}{ \sqrt{2}}\left(v +
\xi_\eta + i \zeta_\eta\right) ; \ \rho^o =  \fr{1}{
\sqrt{2}}\left( u + \xi_\rho + i \zeta_\rho\right);\ \chi^o =
\fr{1}{ \sqrt{2}}\left( w + \xi_\chi + i \zeta_\chi\right).
\label{exp1} \ee
 For the  prime neutral fields which do
not have VEV, we get analogously: \be \eta'^{o} = \fr{1}{
\sqrt{2}}\left( \xi'_\eta + i \zeta'_\eta\right) ; \ \chi'^{o} =
\fr{1}{ \sqrt{2}}\left( \xi'_\chi + i \zeta'_\chi\right).
\label{exp2} \ee

Requiring that in the shifted potential $V$, the linear terms in
fields must be absent, we get  in the tree level approximation,
 the following constraint equations:
\bea \mu^2_1 +  \lambda_1 v^2 +\frac 1 2 \lambda_4 u^2 +\frac 1 2
\lambda_5 w^2
 & = & 0, \nonumber\\
\mu^2_2 +  \lambda_2 u^2 + \frac 1 2 \lambda_4 v^2 +\frac 1 2
\lambda_6 w^2
 & = & 0,
\label{cont}\\
\mu^2_3 +  \lambda_3  w^2 +\frac 1 2  \lambda_5 v^2 + \frac 1 2
\lambda_6 u^2
 & = & 0\nn .
\eea

\hs Since dark matter has to be neutral, then we consider
 only neutral Higgs
sector. In the $\xi_\eta, \xi_\rho, \xi_\chi, \xi'_\eta,
\xi'_\chi$ basis the square mass matrix, after imposing of the
constraints ~(\ref{cont}), has a quasi-diagonal form as follows:

\be M^2_H = \left( \begin{array}{cc}
M^2_{3H}& 0\\
0 & M^2_{2H} \end{array} \right), \ee where \be M^2_{3H} = \fr 1 2
\left( \begin{array}{ccc}
 2 \la_1 v^2  & \la_4 v u   &
  \la_5 v w   \\
  \la_4 v u   &  2 \la_2 u^2 &
  \la_6 u w \\
  \la_5 v w  &  \la_6 u w  &
 2 \la_3 w^2  \end{array} \right),
\label{mat1}
\end{equation}
and
\begin{equation}
M^2_{2H} = \left( \fr{\la_8}{4} +  \la_{10} \right)\Biggl(
\begin{array}{cc}
  w^2    & v w \\
 v w & v^2 \end{array} \Biggr).
\label{mat2}
\end{equation}

The above mass matrix shows that the prime fields mix themselves
but do not mix  with others. In the limit \be \la_1v,\ \la_2 u,\
\la_4 u \ll \la_5 w,\ \la_6 w, \label{lm1} \ee we obtained
physical eigenstates  $H_1(x)$ and   $\si(x)$ \be \Biggl(
\begin{array}{c}
  H_1(x)\\
 \si(x) \end{array} \Biggr)
= \frac{1}{(\la_5^2 v^2 + \la_6^2 u^2)^{1/2}} \Biggl(
\begin{array}{cc}
  \la_6 u   & -\la_5 v \\
 \la_5 v & \la_6 u \end{array} \Biggr)
\Biggl( \begin{array}{c}
  \xi_\eta\\
 \xi_\rho \end{array} \Biggr),
\label{ct17} \ee with masses~\cite{l97} \bea m^2_{H_1}& \approx&
\frac{v^2}{4\la_6}( 2 \la_1 \la_6 -  \la_4 \la_5) \approx
 \frac{u^2}{4\la_5}(
2 \la_2 \la_5 -  \la_4 \la_6), \label{rt1}
\\
m^2_{\si} &\approx&  \fr 1 2 \la_1 v^2 +  \frac{\la_4 \la_6
u^2}{4\la_5} \approx
 \fr 1 2 \la_2 u^2 +  \frac{\la_4
\la_5 v^2}{4\la_6}. \label{rt2} \eea

Eqs.~(\ref{rt1}) and~(\ref{rt2}) also give us relations among
coupling constants and VEVs. Another  massive physical state $H_3$
with mass:
\begin{equation}
  m^2_{H_3} \approx -  \la_3 w^2 .
\label{mass1}
\end{equation}
The scalar $\si(x)$ is the one that we can identify with the SM
Higgs boson~\cite{l97}.

\hs In the approximation $w\gg v$,
 mass matrix $M^2_{2H}$
gives us one Goldstone $\xi'_\chi$ and one physical massive field
$\xi'_\eta$ with mass
\begin{equation}
m^2_{\xi'_\eta} =  -\left( \fr{\la_8}{4} +  \la_{10}\right) w^2.
\label{mass2}
\end{equation}

\hs In  the pseudoscalar  sector, we have three Goldstone  bosons
which can be identified as follows: $ G_2 \equiv \zeta_\eta,\ G_3
\equiv  \zeta_\rho,\ G_4 \equiv  \zeta_\chi$ and in the
$\zeta'^o_\eta, \zeta'^o_\chi$ basis
\begin{equation}
M^2_{2A} = \left( \fr{\la_8}{4} +  \la_{10} \right)\Biggl(
\begin{array}{cc}
  w^2    & v w \\
 v w & v^2 \end{array} \Biggr).
\label{mat3}
\end{equation}
We easily  get one Goldstone  $G'_5$ and one massive pseudoscalar
boson $\zeta'_\eta$ with mass
\begin{equation}
m^2_{\zeta'_\eta} =  -\left( \fr{\la_8}{4} +  \la_{10}\right) w^2.
\end{equation}
It is to be emphasized that, both $\xi'_\eta$ and $\zeta'_\eta$
 are in  an  singlet of the $SU(2)$. Therefore they do not interact
with the SM gauge bosons $W^\pm, Z^0\ \mbox{and}\ \ga$. Unlike the
3-3-1 model considered in~\cite{ft}, here we have two fields which
can be considered as dark matter.

\hs To get the interaction of dark matter to the SM Higgs boson,
we consider the following relevant parts

\bea L_{int}(\si, \zeta_\eta) &=& \fr 1 4 \la_1\left[ v^2 + 2v
\xi_\eta + \xi_\eta^2 + \zeta_\eta^2
+ \xi_\eta^{'2} + \zeta_\eta^{'2} +2\eta^+\eta^- \right]^2 \nn\\
&+& \fr 1 4 \la_4\left[ v^2 + 2v \xi_\eta + \xi_\eta^2 +
\zeta_\eta^2
+ \xi_\eta^{'2} + \zeta_\eta^{'2} +2\eta^+\eta^- \right]\nn\\
&\times&\left[ u^2 + 2u \xi_\rho + \xi_\rho^2 + \zeta_\rho^2 +
2\rho^+\rho^- + 2\rho^{'-}\rho^{'+}\right] \label{shtt} \eea

Substituting (\ref{ct17}) we get  couplings of SIDM with the SM
Higgs boson $\si$ \bea L(\si, \zeta_\eta)&=&\left[
\fr{\si(x)}{\sqrt{\la_5^2 v^2+\la_6^2 u^2}} \left(\la_1 \la_5 v^2
+ \fr{\la_4 \la_6}{2}u^2\right) + \fr{H_1(x)\si(x)}{ (\la_5^2
v^2+\la_6^2 u^2)}\left(\la_1 -\fr{\la_4}{2}\right)
\la_5 \la_6 uv \right.\nn\\
&+& \left. \fr{\si^2(x)}{2(\la_5^2 v^2+\la_6^2 u^2)} \left(
\la_5^2 v^2 +\fr{\la_6^2}{2} u^2\right)\right]\left(\xi_\eta^{'2}
+\zeta_\eta^{'2}\right). \label{sdm} \eea

From Yukawa couplings, we see that our candidates do not interact
with ordinary leptons and quarks~\cite{l96}.
\begin{eqnarray}
{\cal L}_{Yuk}^{\eta}&=&\la_{3a}\bar{Q}_{3L}u_{aR}\eta+
\la_{4ia}\bar{Q}_{iL}d_{aR}\eta^{*}+\mbox{h.c.}\nonumber\\
&=&\la_{3a}(\bar{u}_{3L}\eta^o+\bar{d}_{3L}\eta^-+\bar{T}_L\eta^{,o})
u_{aR}+\la_{4ia}(\bar{d}_{iL}\eta^{o*}-\bar{u}_{iL}\eta^+
+\bar{D}_{iL}\eta^{,o*})d_{aR}+\mbox{h.c.}\nonumber
\end{eqnarray}

We see that the candidates for dark matter in this model have not
couplings with all the SM particles except for the Higgs boson.

\hs For stability of  DM, we have to put mass of the SM Higgs
boson is twice bigger  mass of the candidate \be
 \fr 1 2 \la_1 v^2 +  \frac{\la_4
\la_6 u^2}{4\la_5} \approx
 \fr 1 2 \la_2 u^2 +  \frac{\la_4
\la_5 v^2}{4\la_6} \ge  -\left( \fr{\la_8}{4} + \la_{10}\right)
w^2. \ee

To avoid the interaction of DM with Goldstone boson, we have \be
\la_1 = \fr{\la_4}{2} \ee

\hs  The {\it wrong} muon decay ($\mu^- \rightarrow e^- \nu_e
{\bar \nu}_\mu$) gives a lower limit for singly charged bilepton $
M_Y   \sim 230 \ \mbox{GeV}$.
 Combining Eqns. (\ref{rhb},
\ref{vw}) with (\ref{mar}) we obtain the following relation: $u \
\sim \ v \ \approx 100 - 200 \ \mbox{GeV} $  and  $w \approx (500
- 1000)$ GeV.

The cross section for $h h \rightarrow h h $ (where $h$ stands for
$\xi'_\eta$ and $\zeta'_\eta$) with quartic interaction is $\si =
\la_1^2/4 \pi m_h^2$. The requirement on the quality
$\si_{el}/(m_h[GeV])$ denoting the ration of the
 DM  elastic cross section to its  mass (measured in GeV)
is that~\cite{ss,mcd,far00}

 \be   2.05 \times 10^3 \ \mbox{GeV}^{-3}
 \leq \fr{\sigma}{m_{h}} \leq  2.57 \times 10^4 \
 \mbox{GeV}^{-3} \label{req} \ee Taking
$\lambda_1=1$  we get  4.7  MeV $\le m_{h} \ \le 23 $ MeV. The
SIDM candidates interact with the SM Higgs boson by strength 0.65
if $\la_5 = \la_6 =1$ and $u = v = 175$ GeV are taken.

%This means that couplings constants of DM candidate $\la_1 \sim
%\la_4$ are much small than that of
% non-dark matter ones.

\hs  Now consider the cosmic density of the $h$ scalar given by
~\cite{ft}: \be \Omega_h=2g(T_\gamma)T_\gamma^3
\fr{m_h\beta}{\rho_cg(T)}, \ee
 where $T_\gamma=2.4\times 10^{-4}$ eV
is the present photon temperature, $g(T_\gamma) = 2$ is the photon
degree of freedom and $\rho_c=7.5 \times10^{-47} h^2$ with $
h=0.71 $, being the critical density of the Universe.  Taking $
m_h = 4.7$ MeV,  we obtain $\Omega_h=0.18$. This means that the
SIDM candidates do not overpopulate the Universe.

\hs Recent analysis~\cite{pires} shows that axions and majorons
can be outcome in the 3-3-1 model. As well as the minimal 3-3-1
model, the considered  model contains the Higgs bosons carried
lepton number (scalar bilepton) and Higgs physics in the 3-3-1
models are much richter than that in the SM.

 \hs In conclusion,  we have shown in this paper that the
3-3-1 model with RH neutrinos provides two Higgs bosons: one is
scalar or $CP$-even and another is pseudoscalar or $CP$- odd
particle having properties of  candidates for dark matter. In
difference with the previous candidate which introduced by hand,
our self-interacting dark matter arises without impose new
properties to satisfy all the criteria.  Scalar dark matter
candidates have been recently investigated in~\cite{bf}. The DM
stability could result from the extreme smallness of its couplings
to ordinary particles, it is also necessary to impose a new
symmetry: the $Z_2$ symmetry.
 Recently, astronomical
observations suggest that 70\% of the total energy of the universe
can be associated to the cosmological constant~\cite{net01}. Thus,
the contribution of an exotic particle to dark matter would be
about 30\%. SIDM in our model do not interact with ordinary SM
particle, exclude with the Higgs boson and estimation has shown
that SIDM should be weak.

\hs   Financial support from Swedish International Development
Cooperation Agency (SIDA) through the Associateship Scheme of the
Abdus Salam International Centre for Theoretical Physics, Trieste,
Italy is acknowledged (H. N. L.). This work was supported in part
by National Council for
Natural Sciences of Vietnam contract No: KT - 04.1.2.\\[0.3cm]

\end{document}